# Software Metrics Evaluation Based on Entropy

R. SELVARANI[1]
T.R.GOPALAKRISHNAN NAIR[2]
MUTHU RAMACHANDRAN[3]
KAMAKSHI PRASAD[4]

[1,2] RIIC, Dayananda Sagar Institutions, Bangalore, India
selvss@yahoo.com, trgnair@yahoo.com
[3] Innovation North: The Faculty of Information and Technology
Leeds Metropolitan University, Leeds LS6 3QS, UK m.ramachandran@leedsmet.ac.uk
[4] J.N.T.U, Hyderabad, kamakshiprasad@yahoo.com

**Abstract.** Software engineering activities in the Industry has come a long way with various improvements brought in various stages of the software development life cycle. The complexity of modern software, the commercial constraints and the expectation for high quality products demand the accurate fault prediction based on OO design metrics in the class level in the early stages of software development. The object oriented class metrics are used as quality predictors in the entire OO software development life cycle even when a highly iterative, incremental model or agile software process is employed. Recent research has shown some of the OO design metrics are useful for predicting fault-proneness of classes. In this paper the empirical validation of a set of metrics proposed by Chidamber and Kemerer is performed to assess their ability in predicting the software quality in terms of fault proneness and degradation. We have also proposed the design complexity of object-oriented software with Weighted Methods per Class metric (WMC-CK metric) expressed in terms of Shannon entropy, and error proneness.

## 1 Introduction

Object-oriented design and development is a popular concept in today's software development environment, object oriented (OO) development has proved its value for systems that must be maintained and modified. OO software development requires a different approach from more traditional functional decomposition and data flow development methods, including the metrics used to evaluate OO software [10]. The concepts of software metrics [6][7][8] are well established, and many metrics relating to product design quality have been developed and used. One approach to controlling software maintenance costs is the utilization of software metrics during the development phase, to help identify potential problem areas in the design. Software design complexity is a highly important factor affecting the cost of software development and maintenance. If we can determine the impact of complexity factors on maintenance effort, we can develop guidelines which will help reduce the costs of maintenance by recognizing troublesome situations early in the development phase. In response to these situations the managers can take appropriate decision to reduce the design complexity of the sytem [2][9]. These guidelines will also help to develop tools that support the maintenance of complex modules, to create suitable documentation that helps the developer to manage the complexity in a better way and to allocate the resources.

This paper presents the empirical evaluation of CK

metrics [1][12] for object oriented design based on measurement theory and ontology. These measures applied in a software system can be used to estimate the cost, to schedule the future projects, to evaluate the productivity impacts of new tools and techniques, to establish productivity trends over time, to improve the quality of the software, to forecast future staffing needs, and to reduce future maintenance requirements.

A method based on information theory has also been proposed for examining software design complexity using one of the widely accepted OO complexity design metrics in the context of empirical complexity threshold criteria to assess system-wide software degradation. We have considered five C++ projects done by different group of students. The analysis showed that components with high design complexities were associated with more maintenance activities than those components with lower class complexities.

## 2 Metric Evaluation Criteria

Metrics are defined by Fenton and Pfleeger in [5] as output of measurements, where measurement is defined as the process by which values are assigned to attribute of entities in the real world in such a way as to describe them according to clearly defined rules. Software metrics are the measures of attributes of a software system [3][17]. Traditional functional decomposition metrics and data analysis design metrics measure the design structure independently. OO metrics treats function and data as a combined, integrated object [13][1]. To evaluate a metric's usefulness as a quantitative measure of software quality, it must be based on the measurement of a software quality attribute. The metrics evaluate the OO concepts such as methods, classes, cohesion, coupling, and inheritance. The metrics focus on internal object structure, external measures of the interactions among entities, measures of the efficiency of an algorithm and the use of machine resources, and the psychological measures that affect a programmer's ability to create, comprehend, modify, and maintain software.

## 3 Empirical Literature on CK Metrics[14]

There are a number of empirical studies on CK metrics [1][2] [11][16] [18][24][30]. The existing empirical studies have been compared and the analysis of their results has been reported by Subramanyam and Krishana [30]. To improve the effectiveness of developer interactions in the study, we have adopted a ground theory (can be defined as a systematic qualitative approach to research methodology where research hypothesis and theories can be formulated based on the data collected, [15][31]) dialogue and structured questionnaire to study the effectiveness of the empirical evaluation. Current empirical studies, most notably by Booch [2] and Subramanyam and Krishnan [30] who outlines four major steps involved in the object oriented design process.

1. Identification of Classes (and objects): The key abstractions in the problem space are identified and labeled as potential classes and objects.
2. Identification of semantics of Classes (and objects): The meaning of classes and objects identified in the previous step is established, this includes the definition of the life cycles of each object from creation to destruction.
3. Identify the relationship between Classes (and objects): Classes and objects interactions, such as patterns of inheritance among and patterns of visibility among objects and classes are identified.
4. Implementation of Classes (and objects): Detailed internal views are constructed, including definitions of methods and their various behaviours.In the existing several design methodologies, the design of class is consistently declared to be the central to the OO paradigm. Since the class deals with the functional requirements of the system, it must occur before system design (mapping object to processors and processes) and program design (reconciling of functionality using the target languages, tools etc.). Given the importance of class design the metrics outlined in this paper specifically are designed to measure the complexity of the design of classes. Weyuker has developed a formal list of properties for software metrics and has evaluated a number of existing metrics using these properties [3]. Of nine properties proposed by Weyuker, the following six properties are widely accepted by the researchers.

**Property 1: Non-Coarseness** Given a class P and a metric μ another class Q can always be found such that: μ(P )_μ(Q). This implies that not every class can have the same value for a metric; otherwise it has lost its value as a measurement.

**Property 2: Non-uniqueness (notion of equivalence)** There can exist distinct classes P and Q, $\mu(P) = \mu(Q)$. This implies that two classes can have the same metric value, i, e., the two classes are equally complex.

**Property 3: Design details are important**
Given two class designs, P and Q, which provide the same functionality, does not imply that $\mu(P) = \mu(Q)$. The specifics of the class must influence the metric value. The intuition behind the property 3 is that even though two class designs perform the same function, the details of the design matter in determining the metric for the class.

**Property 4: Monotonicity**
For all classes P and Q, the following must hold: $\mu(P)^2 \mu(P + Q)$ and $\mu(Q)^2 \mu(P + Q)$ Where $P + Q$ implies combination of P and Q. This implies that the metrics for the combination of two classes can never be less than the metric for either of the component classes.

**Property 5: Non equivalence of interaction**
$\exists P$, $\exists Q$, $\exists R$, such that: $\mu(P) = \mu(Q)$ does not imply that and $\mu(Q + R)$. This suggests that the interaction between Q and R can be different than interaction between Q and resulting in different complexity values for $P + Q$ and $Q + R$.

**Property 6: Interaction increases complexity**
$\exists P$, $\exists Q$ such that: $\mu(P)\mu(Q) < \mu(P + Q)$. The principle behind this property is that when two classes are combined, the interaction between classes can increase the complexity metric value.

### 3.1 OO-Specific Metrics:

The OO design metrics are primarily applied to the concepts of classes, coupling, and inheritance. Predicting design defects can save cost enormously. CK suite of metrics has been successfully applied in identifying design defects early during the design process. The summary of CK design metrics are described as follow:

**Weighted Methods per Class (WMC)**
It is a class level metric. A class is a template from which objects can be created. This set of objects shares a common structure and a common behaviour manifested by the set of methods. The WMC is a count of the methods implemented within a class or the sum of the complexities of the methods (method complexity is measured by cyclomatic complexity). The number of methods and the complexity of the methods involved is a predictor of how much time and effort is required to develop and maintain the class. The larger the number of methods in a class, the greater the potential impact on children, since children inherit all of the methods defined in a class. Classes with large numbers of methods are likely to be more application specific, limiting the possibility of reuse. This metric measures the understandability, reusability and maintainability [1][4][5][6][8]. WMC is a good indicator for implementation and test effort.

**Response for a Class (RFC):**
RFC looks at methods and messages within a class. A message is a request that an object makes of another object to perform an operation. The operation executed as a result of receiving a message is called a method The RFC is the set of all methods (internal, external) that can be invoked in response to a message sent to an object of the class or by some method in the class. This metric uses a number of methods to review a combination of a class's complexity and the amount of communication with other classes. If a large number of methods can be invoked in response to a message, testing and debugging the class requires a greater understanding on the part of the tester. A worst-case value for possible responses assists in the appropriate allocation of testing time. This metric evaluates the system design as well as the usability and testability.

As RFC is directly related to complexity, the ability to test, debug and maintain a class increase with an increase in RFC. In the calculation of RFC, inherited methods count, but overridden methods do not. This makes sense, as only one method of a particular signature is available to an object of the class. Also, only one level of depth is counted for remote method invocations.

**Lack of Cohesion of Methods (LCOM)**
Cohesion is the extension of information hiding[5]. Degree to which methods within a class are related to one another and work together to provide well-bounded behaviour. Effective OO designs maximize cohesion because they promote encapsulation. LCOM uses data input variables or attributes to measure the degree of similarity between methods. Any measure of method separateness helps identify flaws in the design of classes. There are two ways to measure cohesion[4]. 1.The percentage of methods that use each data field in a class can be calculated and the average of the percentages can be subtracted from 100 which indicate the level of cohesion. If the percentage is low, the cohesion will be more and if it is high then there will be low cohesion. 2. The count of disjoint sets at from the intersection of the sets of attributes used by the methods also will indicate the level of cohesion. For a good cohesion and less complexity, the class subdivision must be well defined. Classes with low cohesion could probably be subdivided into two or more subclasses with increased cohesion. Any measure of disparateness of methods helps identify flaws in the design of classes. It is a direct indicator of design complexity and reusability.

**Coupling Between Object Classes (CBO)**
Coupling is a measure of the strength of association established by a connection from one entity to another [4]. Classes (objects) are said to be coupled when a message is passed between objects, when methods declared in one class use methods or attributes from the other classes. Tight coupling between super classes and their subclasses is introduced by inheritance. For a good OO design balance between coupling and inheritance is required. CBO is a count of the number of other classes to which a class is coupled [4]. It is measured by counting the number of distinct non inheritance-related class hierarchies on which a class depends. Excessive coupling is detrimental to modular design and prevents reuse. In order to improve modularity and promote encapsulation, inter-object class couples should be kept to a minimum. The larger the number of couples, the higher the sensitivity to changes in other parts of the design; maintenance is therefore more difficult. The higher the inter-object class coupling, the complexity will be increased and more rigorous testing is needed. Complexity can be reduced by designing systems with the weakest possible coupling between modules. This improves modularity and promotes encapsulation [4]. CBO evaluates efficiency and reusability [1][2][3][4][5][6][8].

**Depth of Inheritance Tree (DIT)**
Inheritance is a type of relationship among classes that enables programmers to reuse previously defined objects, including variables and operators [5]. Deep inheritance hierarchies can lead to code fragility with increased complexity and behavioral unpredictability. The depth of inheritance hierarchy is the number of classes (nodes) connected to the main class (root of the tree). The deeper a class within the hierarchy, the greater the number of methods it is likely to inherit, making it more complex to predict its behavior. Deeper trees constitute greater design complexity, since more methods and classes are involved, but the greater the potential for reuse of inherited methods. A support metric for DIT is the number of methods inherited. This metric primarily evaluates efficiency and reuse but also relates to understandability and testability [1][2][3][4][5][6][8].

**Number of Children (NOC)**
For a given class, the number of classes that inherit from it is referred to by the metric Number of Children (number of child classes) [5]. The greater the number of children, the greater the reuse and likelihood of improper parent abstraction, and it may be an indication of sub classing misuse. If a class has a large number of children, it may require more testing of the methods of that class, thus increase the testing time. This metric evaluates efficiency, reusability, and testability of the design of the system. It is an indicator of the potential influence a class can have on the design and on the system [1][4].

## 4 Software Metrics and Entropy Concept

The distinction between reversible and irreversible process was first introduced in thermodynamics through the concept of 'entropy' [22][27]. In the modern context, the formulation of entropy is fundamental for understanding thermodynamic aspects of self organization evolution of order and life that we see in Nature. When a system is isolated, energy increase will be zero. In this case the entropy of the system will continue to increase due to irreversible processes and reach the maximum possible value. This is the state of the thermodynamic equilibrium. In the state of equilibrium, all irreversible process cease. When a system begins to exchange entropy with the exterior then, in general it is driven away from the equilibrium, and the entropy producing the irreversible process begins to operate. This 'state of disorder' is characterized by the amount of disordered energy and its temperature level. Here we have to highlight the following facts as a summary of entropy.

- The entropy of a system is a measure of the amount of molecular disorder within the system.
- A system can only generate but not destroy the entropy.
- The entropy of the system can be increased or decreased by energy transports across the boundary.

The energy sources in the universe were rated on entropy/usefulness scale from zero entropy. The low entropy energy is useful. The use of entropy as a measure of information content of software systems that as led to its use in measuring the code complexity of functionally developed software products. The metric is computed using information available in class definitions. The correlation study used the final versions of class definition. The high degree of positive relationship between entropy based class definition measure and the design complexity measure of class implementation complexity verify that the new entropy measure computed from class definitions can be used as a predictive measure for class implementation complexities provided the class definitions do not change significantly during the implementation. Current studies on entropy [29][28] have been applied mainly to measure the code complexity measures. Our aim in this research is to apply the concept entropy measures for analysis and predict design defects based on grounded empirical analysis which is a structured and interactive approach to user dialogue for collective data based on sociological study. This involves observing how software engineers develop their software and their work environment in which the actual software has been developed. We believe this will have a direct impact on the quality of the software that has been produced. The class complexity related to number of methods in a class is one of the fundamental measures of the 'goodness' of a software design. The most accepted widely studied WMC metric from CK metric suites plays as an important measure for system understandability, testability, and maintainability. This design metrics is a good predictor of time and effort requirement to develop and maintain the class, but when it is associated with entropy metric, it gives an insight about the design degradation or disorder of the system and recommends for redesigning of the system in the early stage itself which in turn reduce the cost of the system.

## 5 Entropy (Information Theory) Based Object Oriented Software System Complexity Measurement

In object-oriented programming, the class complexity measures information flows in a class based on the information passing relationship among member data and member functions. The inter-object complexity for a program measures information flows between objects. Total program complexity is measured by class complexity and inter-object complexity. The term 'software entropy' has been defined to mean that software declines in quality, maintainability and understandability through its lifetime. Here Shannon's entropy equation is used to establish a measure of OO software degradation that is easy to use and interpret. WMC (weighted method per class), a well-established CK metrics is used to asses this criteria. WMC thresholds are the basis for our metric measurement. We have used the threshold criteria for WMC published by Rosenberg, et al. Software Assurance Technology Center (SATC), NASA Goddard Space Flight Center, in 1998 [19]. These thresholds were based on their experiences at NASA with OO projects. It is shown in Table 1, and will be used without modification in this application. Table 1 gives the threshold criteria and interpretation of risk based on NASA-SATC guidelines[19]. The use of these thresholds in industry allows software managers to make judgments about the class complexity of their software in terms of effort required for testing the system and the level of confidence required in software deployment.

**Table 1:** CK-WMC Threshold- NASA-SATC Data

| System Category | CK-W MC Threshold (x) | Risk Interpretation |
|---|---|---|
| 1 | $1 \leq x \leq 20$ | Good values of class complexity. |
| 2 | $20 \leq x \leq 100$ | Moderate high values of complexity. |
| 3 | $x > 100$ | High class complexity, cause for investigation |

### 5.1 Properties of Shannon's Entropy:

The Shannon entropy, $H_n$, is defined as:

$$H_n(F) = -\sum_{k=1} F_k = 1(a > 1) \quad (1)$$

$$F_k > 0(k = 1, ...., a) \text{ aad } \sum_{k=1} F_k = 1(a > 1) \quad (2)$$

Where,
H System=System Complexity Entropy.
k=Integer value 1, 2, ...$j$ representing each of the categories considered.
$F_k$=Total number of classes that are in category F .
N =Total number of system cases (equal to the sum of all the $F_k$s).

Because a logarithm to the base 2 is used, the resulting unit of information is called the bit (a contraction of binary unit). The Shannon entropy satisfies many desirable properties. The following properties of the selected mathematical approach are more suitable for this application [21].

**1. Non negativity:** Information about an experiment makes no one more ignorant than he was before [28][29].

$$H_n(F) > 0 \quad (3)$$

**2. Symmetry:** The amount of information is invariant under a change in the order of events.

$$H_n(F) = H_n(p_{k(1)}, p_{k(2)}, .....p_{k(n)}) \quad (4)$$

Where k is an arbitrary permutation on $\{1, 2....a\}$

**3. Normality:** A "simple alternative", which in this case is an experiment with two outcomes of equal probability 0.5, promises one unit of information.

$$H_2(0.5, 0.5) = 1 \quad (5)$$

**4. Expansibility:** Additional outcomes with zero probability do not change the uncertainty of the outcome of an experiment.

$$H_n(p) = H_{n+1}(F_1, F_2, .....F_n, 0) \quad (6)$$

**5. Decisivity:** There is no uncertainty in an experiment with two outcomes, one of them is the Non-negativity of probability 1, the other is of probability 0.

$$H_2(1, 0) = 0 \quad (7)$$

**6. Additivity:** The information, expected from two independent experiments, is the sum of the information expected from the individual experiments.

$$H_{nm}(F * Q) = H_n(F) + H_m(Q) \quad (8)$$

**7. Subadditivity:** The information, expected from two experiments, is not greater than the sum of the information expected from the individual experiments.

$$H_{nm}(F * Q) \leq H_n(F) + H_m(Q) \quad (9)$$

**8. Maximality:** The entropy is greatest when all admissible outcomes have equal probabilities.

$$H_n(F) \leq H_n(1/a, 1/a, 1/a, .....1/a) \quad (10)$$

### 5.2 Measures of information and their characterizations

The concept of entropy, as a measure of information, is fundamental in information theory. The entropy of an experiment has dual interpretations. It can be considered both as a measure of the uncertainty that prevailed before the experiment was accomplished and as a measure of the information expected from an experiment [20]. An experiment might be an information source emitting a sequence of symbols (i.e., a message) $M = \{s1, s2, s3, ..., sa\}$, where successive symbols are selected according to some fixed probability law, with which the symbols occur $F = (p1, p2, ..., pa)$ [22][23].

In this paper the uncertainty measure that prevailed before the experiment is performed. The maximum entropy is achieved when $S_i = S_{i+1} = S_{i+2} = S_{i+3}$ , or when all the classes are evenly distributed. Shannon's equation "dampens" the effect of a few very highly complex methods to skew the overall complexity of the system. This is because the equation limits the contribution of the entropy score from each category to the overall (system) entropy score.

### 5.3 The Shannon's entropy relationship

Shannon's Entropy equation[26] provides a way to estimate the average minimum number of symbols based on the frequency of the symbols. By treating the software system as an information source, the function calls or method invocation in object oriented systems resemble the emission of symbols from an information source. Thus the probabilities required for computing the entropy are obtained using an empirical distribution or function calls or method invocations.

## 6 Experimental Analysis

If we treat a software system as an information source then the symbols emitted from the system can be the operators within a program, where operators are a special symbol, a reserved word, or a function call [23]. Another technique can be based on data flow relationships [24]. The technique adopted here considers the function calls in procedural programming as the symbols emitted from a software system (or module). In object oriented programming, we replace function calls with method invocations. The rationale behind this choice is that performing calls to different functions resembles emitting a message of many symbols particular to the considered module. The complexity of the design in object oriented system is the weighted method per class.

The probabilities are obtained using an empirical distribution of the function calls. The WMC metric measurement by NASA SATC is based on the number of distinct functions or modules in a class and the complexity is the message transfer between the modules in the class.

The WMC complexity measurement is done by considering the different summations, in the definitions of entropies, over the number of distinct functions or modules in a class. In this design metric, there is no possibility of 0 modules in any of the classes, hence the WMC metric recommended by NASA-SATC starts from 1. The information will be zero if there are no functional calls in a module. We have considered the following five different Java projects by different teams of stu-dents as examples to demonstrate the application of our technique to understand the disorderliness of the project.

This model is used to predict the disorderliness associated with the system in the class level. Table 2 depicts the program metrics obtained by analyzing the projects with automated tool Understand Java. The total number of classes in each project as shown in Figure 1. is divided in to samples according to the algorithm shown in table 1.

The measures calculated are Shannon generalized entropies as given by equation (1) and the results are consistent. As stated by the designer in the program's documentation: "The only algorithms at all difficult are those for parsing, which are rather ad hoc but apparently correct" [25]. This fact is identified by this information measure, which have the highest value for the module of higher design complexity. The next highest value was appropriately given to the module of comparatively lesser design complexity. If we check the rest of the classes, it is clear that the information content measures give meaningful and intuitive results.

**Table 2:** Project Metrices

| Project Metric | P1 | P2 | P3 | P4 | P5 |
|---|---|---|---|---|---|
| Classes: | 37 | 46 | 120 | 139 | 148 |
| Files: | 35 | 34 | 56 | 65 | 90 |
| Library Units: | 209 | 234 | 267 | 168 | 289 |
| Lines Blank | 788 | 675 | 1253 | 1569 | 2378 |
| Lines Code: | 3258 | 8567 | 8450 | 11236 | 12564 |
| Lines Comment: | 2759 | 7498 | 7456 | 9606 | 10997 |
| Lines Inactive: | 0 | 0 | 0 | 0 | 0 |
| Executable Statements: | 1604 | 5078 | 4589 | 6752 | 7629 |
| Declarative Statements: | 791 | 1126 | 569 | 2319 | 2746 |
| Ratio Comment/ Code: | 0.85 | 0.87 | 0.88 | 0.89 | 0.88 |

The goal of object oriented design, is "to design the classes identified during the analysis phase and the user interface". In this design model, the system architecture may have a large number of simple classes, rather than a small number of complex classes for better reusability and maintainability, which in turn displays lesser design complexity. Figure 1 depicts the class distribution among the sample projects of our study. It is observed that the project with larger number of classes is comparatively less prone to degradation, because the entropy $\alpha$=0.59. The entropy of a software system is a class of metrics to assess the degree of disorderliness in a software system structure. Entropy covers all the components of a software system at different abstraction levels, as well as the traceability and relationships among them. It is a direct measure for design complexity and quality of the system.

**Table 3:** Java projects entropy degradation- WMC

| Project | Total Classes | S1 | S2 | S3 | WMC Entropy $\alpha \leq 1$ | N*(WMC Entropy) |
|---|---|---|---|---|---|---|
| P1 | 38 | 34 | 3 | 1 | 0.5<br>46781 | 20.7<br>77678 |
| P2 | 46 | 38 | 6 | 2 | 0.8<br>07802 | 37.1<br>58811 |
| P3 | 120 | 105 | 12 | 3 | 0.6<br>33912 | 76.0<br>694421 |
| P4 | 139 | 126 | 7 | 6 | 0.5<br>41312 | 75.2<br>433682 |
| P5 | 148 | 132 | 11 | 4 | 0.5<br>91036 | 87.4<br>733282 |

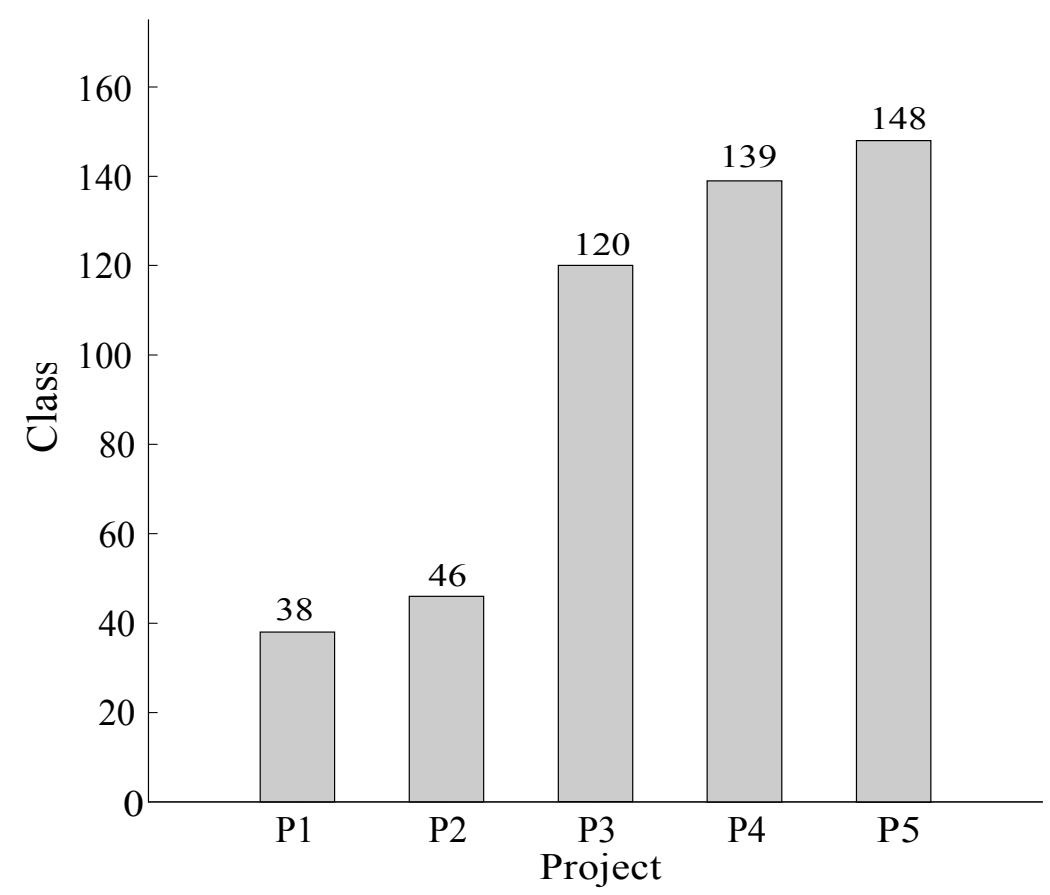

**Figure 1:** Project-Class Distribution

Table 3. depicts the result of application of Shannon entropy equation to verify the utility of complexity metrics for predicting the complexity of initial OO classes. The NASA SATC WMC threshold criteria are used to form the sample set of classes in each project. In this analysis it is observed as the degradation level of project 2 is higher than other projects. Figure 2 depicts the complexity levels of sampled project P1 to P5.

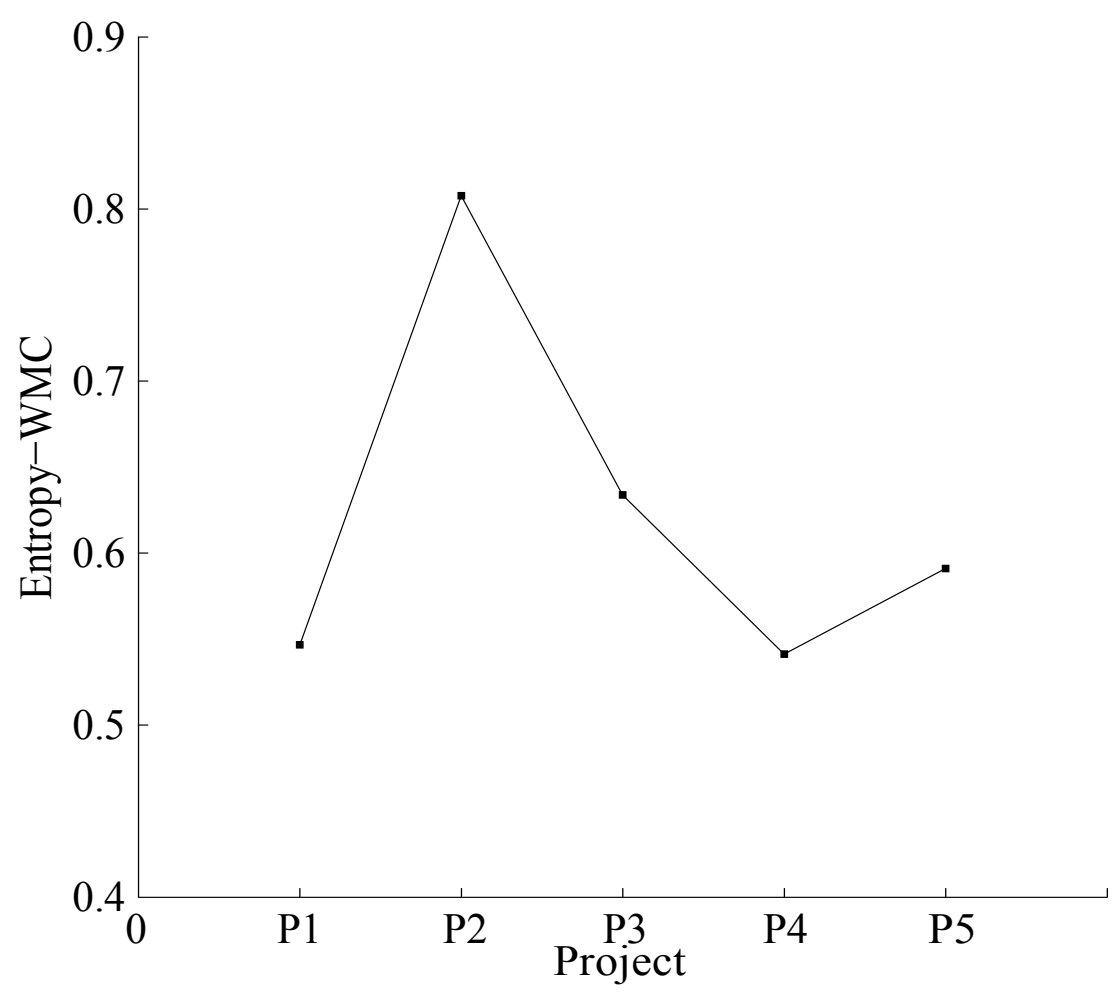

**Figure 2:** Project Complexity

## 7 Conclusion And Future Work

The benefits of object-oriented programming are the resulting simplicity and understandability of the problem through the use of abstraction. However, even OO software is not immune to the effects of brittleness, or degradation. We believe that this entropy degradation metric with OO design metrics thresholds may be useful in evaluating OO software, specifically large Java and C++ systems. This metric may be of most value in programming environments where legacy code is being reengineered into object-oriented programs. We have developed a model based on Shanon's entropy equation (eqn-1) with their mathematical properties (non negativity, symmetry, normality, Expansibility, decisivity and maximality, additivity and subadditivity) to measure the design complexity of the projects with the CK-WMC metric using the variations of the WMC metric and widely accepted threshold values for interpreting the complexity. The measure based on the Chidamber and Kemerer version of WMC, where a complexity score of '1' is assigned to each method in a class showed the most promise at being a good indicator of system degradation. The group of classes with higher entropy scores are more prone for degradation it is extremely difficult in assessing the module independency in a software system. Hence the complexity score the 'Shannon entropy' is of degree $\alpha \leq 1$.

The probability for computing the entropies are obtained using the empirical distribution of the methods in a class. The Shannon entropy is more consistent for different values of WMC metric. As $\alpha$ increase, the measure becomes coarse and indicates the high degradation possibilities of the Object oriented software system. The NASA/Rosenberg threshold risk criteria provided the best correlation to system degradation, because of the grouping of the classes into three categories according to the metric criteria. Software measurement has been a successful approach in evaluating and predicting process capability through personnel performance. Future research includes the system inhomogenity measurement with complete set of CK metric suite and also assessing the performance of various teams involved in developing the software products.

The entropy model generated here have produced results which are useful and are capable of providing effective guidelines during the design time to the design architect to reduce the system entropy by appropriately adjusting design metrics. This approach is effective, useful and promising towards developing a better quality, cost effective software product.